\documentclass[conference]{IEEEtran}

\usepackage{amsmath} 
\usepackage{graphicx}
\usepackage{amsfonts}
\usepackage{float}
\usepackage{color}
\usepackage{cite}
\usepackage{epstopdf}
\usepackage{booktabs,subcaption,dcolumn}
\usepackage{hyperref}
\usepackage{array}
\usepackage{multirow}
\usepackage{multicol}
\usepackage{caption}
\usepackage{caption}
\newcolumntype{P}[1]{>{\centering\arraybackslash}p{#1}}

\pdfoutput=1

\title{Angular and Temporal Correlation of V2X Channels Across Sub-6 GHz and mmWave Bands}

\author{Chethan Kumar Anjinappa and Ismail Guvenc \\
\IEEEauthorblockA{Department of Electrical and Computer Engineering, North Carolina State University, Raleigh, NC}
Email: {\tt \{canjina,iguvenc\}@ncsu.edu}

\vspace{-0.2cm}
}

\begin{document}

\maketitle
\thispagestyle{empty}
%
\pagestyle{empty}

\begin{abstract}
5G millimeter wave (mmWave) technology is envisioned to be an integral part of next-generation vehicle-to-everything (V2X) networks and autonomous vehicles due to its broad bandwidth, wide field of view sensing, and precise localization capabilities. The reliability of mmWave links may be compromised due to difficulties in beam alignment for mobile channels and due to blocking effects between a mmWave transmitter and a receiver. To address such challenges, out-of-band information from  sub-6 GHz channels can be utilized for predicting the temporal and angular channel characteristics in mmWave bands, which necessitates a good understanding of how propagation characteristics are coupled across different bands. In this paper, we use ray tracing simulations to characterize the angular and temporal correlation across a wide range of propagation frequencies for V2X channels ranging from 900~MHz up to 73~GHz, for a vehicle maintaining line-of-sight (LOS) and non-LOS (NLOS) beams with a transmitter in an urban environment. Our results shed light on increasing sparsity behavior of propagation channels with increasing frequency, and highlight the strong temporal/angular correlation among 5.9~GHz and 28~GHz bands especially for LOS channels. 
\end{abstract}


\section{Introduction}
Next generation intelligent transportation systems (ITSs) 
aim to enable safer, more efficient, and informed road experience for all road users, by applying wireless and IoT technologies to allow for vehicle-to-everything (V2X) communication across all elements of road transportation~\cite{ITS_Survey}. Existing technologies for ITS such as 4G LTE and dedicated short range communication (DSRC) fall short of meeting all requirements of next generation ITS communications that require high data rates, precise ranging/sensing capability, and ultra-reliable low latency communications (URLLC). 
5G technology involving millimeter wave (mmWave) communications is expected to enable several new usecases for autonomous driving~\cite{mc2015independent} and V2X communications, such as platooning capabilities, collusion/queue/curve warning, 
intersections without traffic lights, and in-vehicle multimedia entertainment systems~\cite{mmWave_V2X_Survey}. 



The mmWave systems will likely be initially deployed in conjunction with lower frequency (sub-6~GHz) \emph{anchor network} to provide wide area control signals~\cite{mmWave_Sub6GHz}. The performance of a vehicular (wireless) communication system critically depends on the accuracy with which the channel is estimated, and the channel estimation process can be a source of significant overhead in establishing mmWave links, especially for vehicular scenarios with high mobility. Since the mmWave networks will co-exist along with the sub-6 GHz communication networks, one can exploit the out-of-band (OOB) information obtained from sub-6 GHz band as a side information to improve channel estimation in mmWave bands~\cite{7218630,choi2016millimeter,7888146,8198818}. It therefore carries critical importance to better understand the behavior of V2X propagation channel across different vehicular bands. 

Use of side information from WiFi bands to improve mmWave beam steering is studied for indoor scenarios in~\cite{7218630}. In~\cite{choi2016millimeter}, it is considered that side information such as position, velocity, size, and heading can be communicated from neighboring vehicles using the DSRC technology to improve beam alignment at mmWave frequencies. In~\cite{7888146}, authors estimate the narrow-band multi-antenna channel at mmWave bands, using the channel estimates at lower, sub-6~GHz bands. In particular, the spatial correlation matrix from a uniform linear array at lower bands is translated to correlation matrix at mmWave bands through interpolation and extrapolation. A better-performing parametric approach which theoretically characterizes high-frequency correlation is also studied, requiring the knowledge of receiver array geometry and the distribution of the direction of arrival. In~\cite{8198818}, channel training overhead reduction due to use of sub-6 GHz side information for beam steering is explored. 

To our best knowledge, the correlation among sub-6~GHz and mmWave signals considering their temporal and angular domain features have not been studied in the literature for vehicular scenarios. In particular, we consider a wide band single-input single-output channel, and study some key multipath statistics of the time-of-arrival (TOA), angle-of-arrival (AOA), and angle-of-departure (AOD) of vehicular channel realizations across a wide range of frequencies, and attempt to capture the channel correlation across different bands. 
To analyze whether channel characteristics in lower bands can be used to predict channel at mmWave frequencies, we set-up a simple model in Remcom Wireless Insite ray tracing simulator~\cite{WirelessInsite}. Subsequently, we study the correlation characteristics across different sub-6 GHz and mmWave bands, ranging from 900~MHz up to 73~GHz, for a vehicle traveling in an urban environment for line-of-sight (LOS) and non-LOS (NLOS) scenarios. 

\section{System Model}\label{System_Model}

Consider the following physical propagation model between the transmitter and receiver through the omni-directional 
channel impulse response (CIR):
\begin{equation}\label{DCIR1}
h(t,\boldsymbol{\tau},\boldsymbol{\psi},\boldsymbol{\theta}) = \sum_{k=1}^{\tilde{K}(t)} \underbrace{\rho_k e^{-j \phi_k}}_\text{AnP term} \underbrace{\delta (\tau - \tau_k)}_\text{TOA term} \underbrace{\delta (\psi - \psi_k)}_\text{AOA term} \underbrace{\delta (\theta - \theta_k)}_\text{AOD term}
\end{equation}
where $t$ is the time at which the channel is excited, $\tau_k$ is the considered time delay, $\psi_k$ and $\theta_k$ are the AOA and AOD associated with the $k^\text{th}$ multi-path component (MPC), respectively. $\rho_k$ is the amplitude which is based on the path loss and shadowing associated with the $k^\text{th}$ MPC and $\phi_k$ is the phase component associated with the $k^\text{th}$ MPC which primarily depends on the time delay and the Doppler shift. All the TOA, AOA, and AOD components are further clustered into the vectors $\boldsymbol{\tau}=[\tau_1,...,\tau_{\tilde{K}(t)}]$, 
$\boldsymbol{\psi}=[\psi_1,...,\psi_{\tilde{K}(t)}]$, and $\boldsymbol{\theta}=[\theta_1,...,\theta_{\tilde{K}(t)}]$, respectively. The amplitude and phase terms are merged under the amplitude and phase (AnP) variable, $\boldsymbol{\alpha}~=[\rho_1 \exp(-j\phi_1),...,\rho_{\tilde{K}(t)} \exp(-j\phi_{\tilde{K}(t)})]$. Intuitively, (\ref{DCIR1}) is the sum of the contributions of discrete MPCs, where, the number of MPCs $\tilde{K}(t)$ may itself be time-varying due to the mobility of the vehicle and the surrounding scatterers.

A main goal in this paper is to investigate the question \emph{``What can we say about the CIR $h(t,\boldsymbol{\tau},\boldsymbol{\psi},\boldsymbol{\theta},f_i)$ at frequency band $f_i$, given that we know the CIR $h(t,\boldsymbol{\tau},\boldsymbol{\psi},\boldsymbol{\theta},f_j)$ at frequency band $f_j$?"}. 
We are primarily interested in the 5.9~GHz DSRC band commonly considered for V2X communication, and the popular 28~GHz mmWave band.
One simple way to look into this problem is to see if we can have one-to-one mapping for the AnP, TOA, AOA and AOD terms from $f_i$ to $f_j$. If there exists a way to translate information from one frequency band to another then it is easy to answer the question. 
However, while the mechanism for RF propagation is the same, their subtle effects are different at different frequencies. For instance, at mmWave frequency, diffraction is not as important, scattering can be higher, MPCs observe higher path loss, and blockage is more significant. 

To quantify the relation between the TOA, AOA, and AOD at $f_i$ and $f_j$, we study their correlation using representative ray tracing simulations for a V2X scenario. The importance of studying correlation across different frequency bands is two-fold. First, correlation (or lack of correlation) between the variables of interest helps us to rule out (or not rule out) the existence of a one-to-one mapping. Second, correlation can be used to make the prediction at $f_i$, provided estimates at $f_j$. If variables of interest are highly related then it can be used to make a more accurate prediction when side information from another band is provided. We revisit the question on side information once we find out the level of correlation across different bands. In the next section, we describe the simulation scenario to study the correlation across different bands.

\begin{figure}[t!]
\centering 
\includegraphics[scale=0.425]{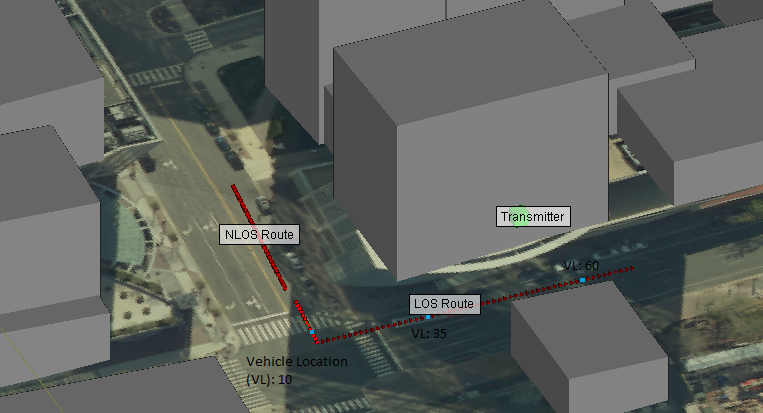}
\caption{Side view: Vehicular ray-tracing scenario in North-Moore Street, Rosslyn, Virginia.}
\label{Ray_Tracing1}\vspace{-0.5cm}
\end{figure}

\section{Ray-tracing Simulations}\label{Sec:RT}


In this work a full 3D ray-tracing (RT) simulator, Wireless Insite~\cite{WirelessInsite}, is used as a tool for the analysis of site-specific radio wave propagation. 
In the past, RT has been reasonably successful at predicting site-specific mmWave propagation~\cite{Comparison_RayTracing}. The mmWave frequencies can be well approximated by ray optics approaches as the diffraction is low and scattering are mostly limited to surface reflections~\cite{Air_Interface_Ray_Tracing_Study}. Hence, with proper approximations RT simulators can accurately model the mmWave and sub-6~GHz channels.
We consider a vehicle-to-infrastructure (V2I) scenario where a moving vehicle communicates with an infrastructure (cellular base station (BS)) as in Fig.~\ref{Ray_Tracing1}. We set~up a vehicular user equipment (UE) (red path) and a BS node (green point). All the nodes use half-wave dipole antennas that are excited with varying frequencies. For simplicity, single input single output (SISO) model is considered in this work, that is, we assume both the transmitter and receiver are equipped with a single antenna. In our future work, we will extend it to the multiple antenna case.

\begin{table}[t!]
\caption{Frequency bands used in ray tracing simulations.}
\begin{tabular}{|P{2.05cm}|P{0.325cm}|P{0.325cm}|P{0.325cm}|P{0.325cm}|P{0.325cm}|P{0.325cm}|P{0.325cm}|P{0.325cm}|}
\hline
\textbf{Band} & \textbf{$f_1$} &  \textbf{$f_2$} &  \textbf{$f_3$} &  {\textbf{$f_4$}} &  \textbf{$f_5$} & {\textbf{$f_6$}} &  \textbf{$f_7$} &  \textbf{$f_8$}\\
\hline
\textbf{Frequency(GHz)}  & 0.9& 1.8 & 2.1 &  {5.0} & 5.9& {28} & 38 & 73\\
\hline
\end{tabular}\label{tab:T_FI}\vspace{-.5cm}
\end{table}

The BS antenna in the RT simulations is placed at a height of 10~m from the ground and transmits at a power level of 0~dBm. The moving UE antenna is placed at a height of 2~m from the ground and takes the indicated route in Fig.~\ref{Ray_Tracing1}; this can be implemented in Wireless Insite using the ``Route" feature. There exists a total of 100 vehicle locations (points) in the route: the first 30 points correspond to the NLOS trajectory and the latter 70 points corresponds to the LOS trajectory. The rationale behind choosing separate LOS and NLOS paths is that the mmWave channel is often a LOS or quasi-LOS channel with few dominant path clusters. 
Thus, it is expected that if there exists correlation, then it will be lower in the NLOS case relative to the LOS case, which will be discussed further in Section~\ref{Simulation_Section}. With these settings, we carry out RT simulations considering the popular sub-6 GHz and mmWave bands shown in Table~\ref{tab:T_FI}.


From RT simulation, we collected a total of 250 MPCs for each vehicle location on the route. We note that out of 250 paths only a few paths have non-negligible energy. In order to limit the number of paths, we employ a thresholding approach. That is, we consider the path to be present if and only if it is above a threshold from the main tap (most dominant path). We refer to this threshold as our ``MPC Threshold (MPCT)". The choice of MPCT depends on the trade-off between an accurate representation of the channel and implementation complexity. Based on the statistics obtained from our experiment, we set MPCT to 40 dB; See section \ref{MPC_Stats} for details. Thus, before we perform any data processing on the collected data, we apply an MPCT of 40 dB on the collected data and process it further. We note that channel parameters extracted from the data can be significantly different when using a different MPCT value. The next section provides the statistics obtained from data collected through RT simulation\footnote{Matlab script and data are available online: \url{https://research.ece.ncsu.edu/mpact/data-management/}}.

\section{Multipath Channel Statistics over sub-6 GHz and mmWave Bands}\label{MPC_Stats}

\begin{figure}[t!]
    \centering
    \begin{subfigure}[t]{0.5\textwidth}
			\centerline{\includegraphics[scale=0.6]{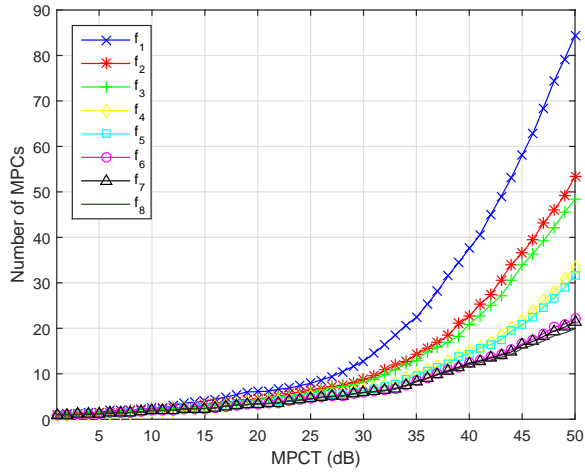}}
            \caption{LOS trajectory.}
    \end{subfigure}%
  
    \begin{subfigure}[t]{0.5\textwidth}
       \centerline{\includegraphics[scale=0.6]{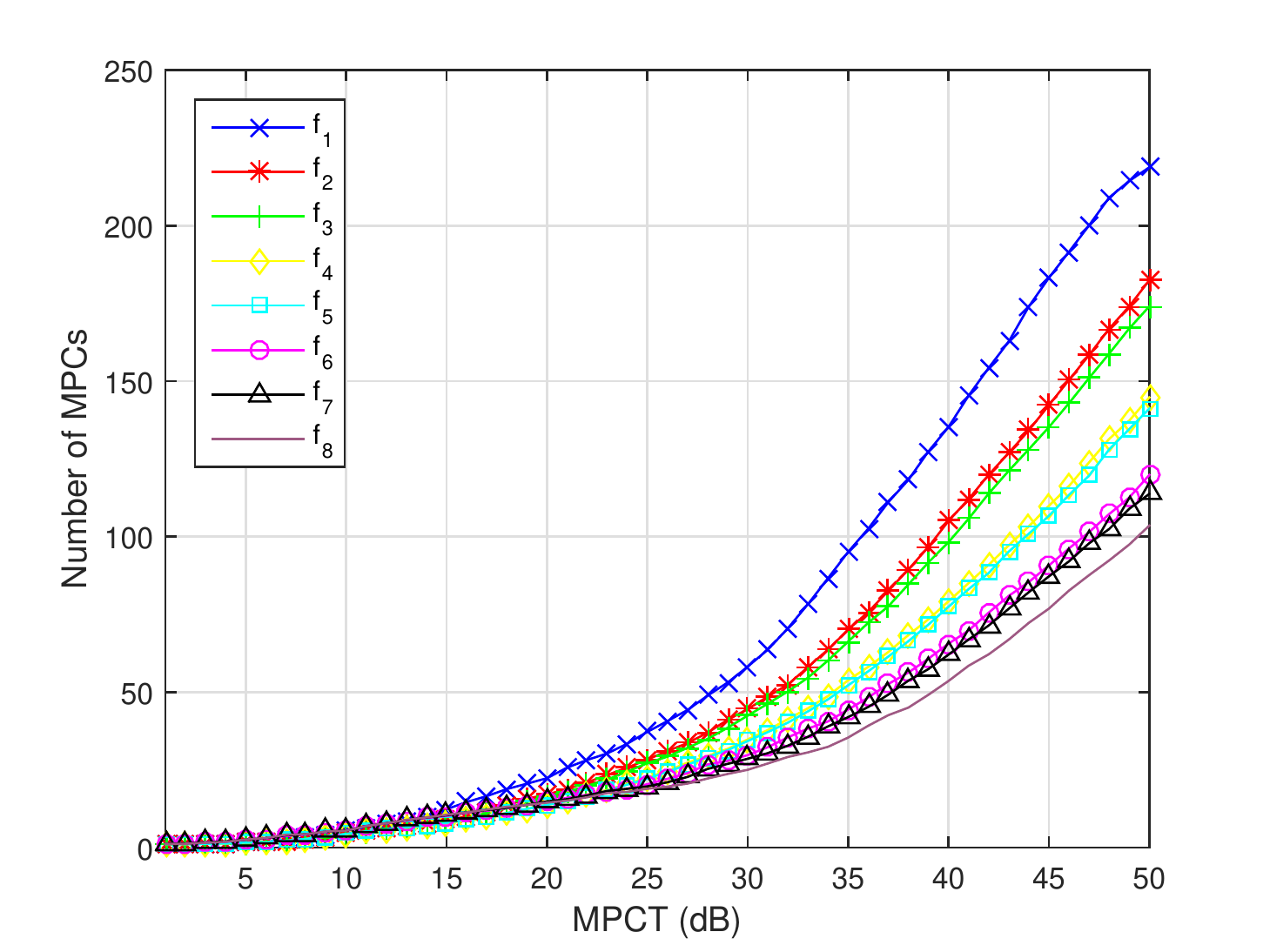}} 
        \caption{NLOS trajectory.}
    \end{subfigure}
    \caption{Number of MPCs as a function of MPCT.}
    \label{MPC_Vs_MPCT}
\end{figure}

Based on the RT simulation framework in Section~\ref{Sec:RT}, in this section we provide some key multipath channel statistics across different frequency bands corresponding to the same vehicular trajectory in Fig.~\ref{Ray_Tracing1}, to better understand how the V2X propagation behavior are coupled across sub-6~GHz and mmWave bands. First, 
Fig.~\ref{MPC_Vs_MPCT} plots the average number of MPCs as a function of MPCT at different frequency bands for both the LOS and NLOS scenarios, respectively. As seen the average number of MPCs monotonically increases with the MPCT threshold. This is because with increase in threshold, more MPCs fall within the threshold range. Also, the average number of MPCs decreases with increase in frequency because of higher path loss with traveled MPC distance for higher frequencies. With increase in MPCT, there will be a considerable mismatch in the number of MPCs at different frequencies.

\begin{figure}[h!]
    \begin{subfigure}[t]{0.5\textwidth}
\centerline{\includegraphics[scale=0.6]{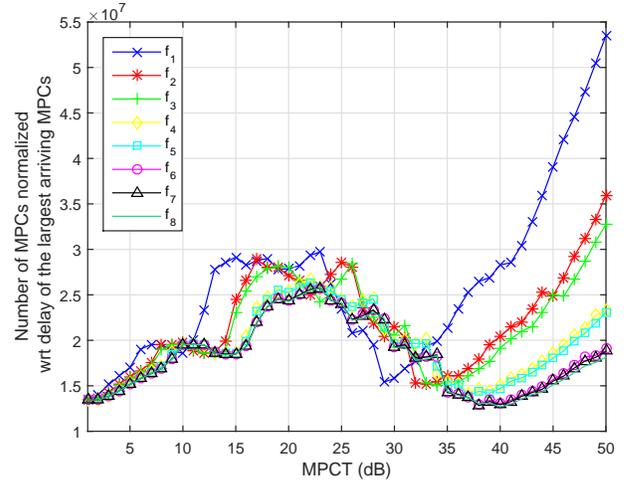}}
	\caption{LOS trajectory.}
\label{NMPC_Vs_MPCT_LOS}
    \end{subfigure}
    \begin{subfigure}[t]{0.5\textwidth}
\centerline{\includegraphics[scale=0.6]{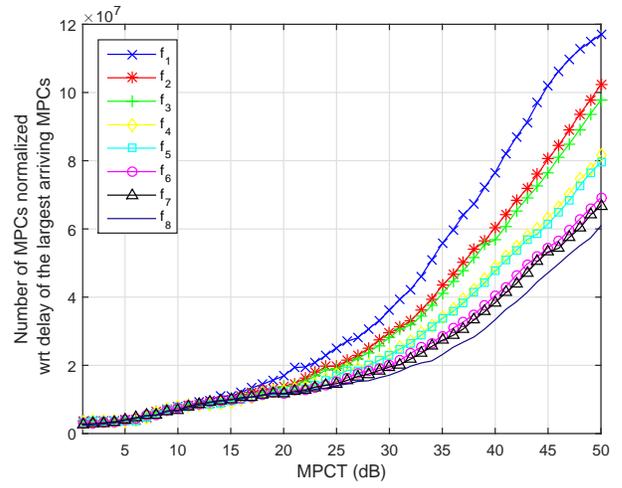}}
\caption{NLOS trajectory.}
\label{NMPC_Vs_MPCT_LOS}
    \end{subfigure}
    	\caption{Average normalized number of MPCs with respect to largest delay of the arriving MPCs as a function of MPCT.}
    \label{NMPC}
\end{figure}

Fig. \ref{NMPC} plots the number of MPCs normalized with respect to the excess delay of the latest arriving MPC as a function of MPCT for both the LOS and NLOS scenarios. 
This normalized metric is expected to be more representative of the \emph{sparsity} of the channel, since it directly captures the time support of the propagation channel. 
Note that the metric decreases with the increase in the maximum delay and increases with the number of MPCs when the largest delay of MPC has already arrived. 
At first glance, Fig.~\ref{NMPC}(a) may seem counter-intuitive, this is because the metric initially increases with the MPCT followed by a downfall and then further increase monotonically beyond a point. The trend is similar at all the frequency bands. The possible explanation for this phenomenon is that at lower MPCT the LOS paths arrive upfront followed by the MPCs clustered around the LOS MPC. These MPCs clustered around the LOS MPC increases the number of MPCs and are often single bounce reflections which slightly increases the largest delay resulting in the slight increase of the metric (increase in the number of MPCs is significant) as seen in Fig.~\ref{NMPC}(b). As we further increase the threshold, later on, NLOS paths are pulled into the picture which results in the increase in the delay by large margin resulting in the reduction of the metric. Beyond a point, the latest arriving MPC with significant power is captured. After this point, the increase in threshold only increases the MPCs which arrive earlier and at lower strength relative to the latest arriving MPC. 

With an MPCT of 40~dB in Fig.~\ref{NMPC}(a), the latest arriving MPC with strong power can be captured; for higher MPCT all the incoming MPCs tend to have a lower delay but a higher path loss because of multiple bounce specular reflections which decreases the signal strength. However, the same phenomenon is missing in the NLOS trajectory in Fig.~\ref{NMPC}(b). 

\begin{figure}[t!]
    \centering
    \begin{subfigure}[t]{0.5\textwidth}
\centerline{\includegraphics[scale=0.6]{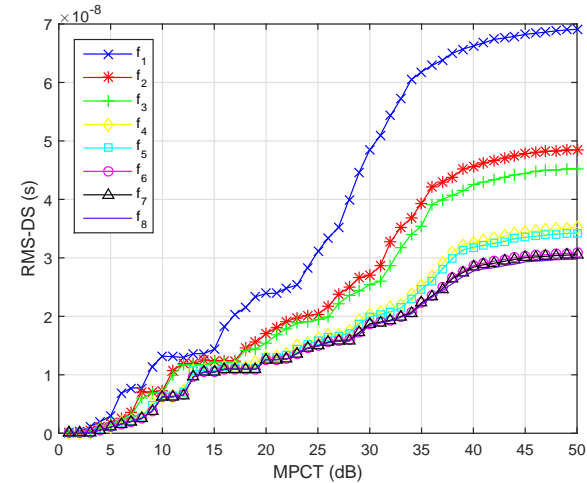}}
            \caption{LOS trajectory.}
    \end{subfigure}%
  
    \begin{subfigure}[t]{0.5\textwidth}
\centerline{\includegraphics[scale=0.6]{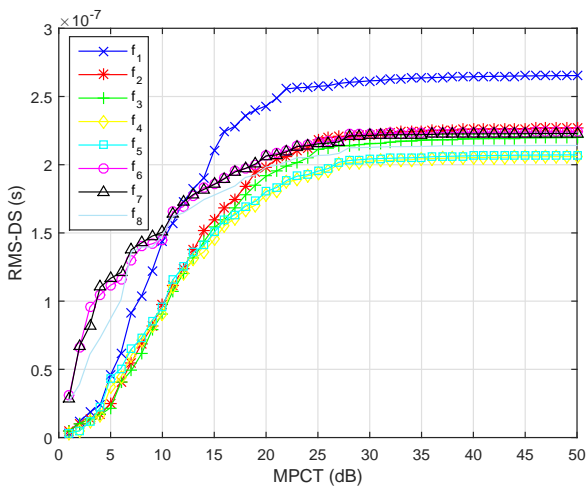}}
        \caption{NLOS trajectory.}
    \end{subfigure}
    \caption{RMS-DS as a function of MPCT.}
    \label{RMS_DS_Vs_MPCT}
\end{figure}

In Fig.~\ref{RMS_DS_Vs_MPCT}, we present the root-mean-square delay spread (RMS-DS) of the MPCs as a function of MPCT captured at different frequency bands for the LOS and NLOS trajectories, respectively. We see that the RMS-DS increases with the increase in MPCT and saturates beyond a point for both the LOS and NLOS trajectory. From the plots, it is evident that setting MPCT to 40 (25) dB would capture the RMS-DS for the LOS (NLOS) scenario significantly in all the frequency bands. For consistency, we set the MPCT to 40~dB in the sequel for both the LOS and NLOS scenarios.

\section{Correlation Across Different Bands}

In order to assess the temporal and angular correlation across different bands, we assume TOA, AOA, and AOD terms as random variables (RVs) and that they are independent. 
Note that in practice, there is some correlation between these RVs, and independence assumption will not usually hold. 
The RVs for correlation are defined for different parameters as follows.

\textbf{Step 1:} MPC thresholding:
\begin{align*}
{\boldsymbol{\alpha}_\text{MPCT}(k)}= 
\begin{cases}
{\alpha_k}~,&  \text{if}~|\alpha_k|\geq\max|\boldsymbol{\alpha}| - \text{MPCT} \\
0~, & {\rm otherwise}~
\end{cases}~,
\end{align*}
where the index $k$ refers to the $k^{\rm th}$ MPC, $\boldsymbol{\alpha}$ corresponds to the AnP vector as defined in Section~\ref{System_Model}, and the vector $\boldsymbol{\alpha}_\text{MPCT}$ contains the MPCs that are within the threshold range. In other words, it contains only the MPCs that have non-negligible energy. We next proceed to place the MPCs in a vector according to the bins in which they fall. That is, we form a vector $\boldsymbol{h^{(x)}}$ of size $N$, where the value of $N$ depends on the domain and the resolution of choice. 
The process of constructing a sparse vector from $\boldsymbol{\alpha}_\text{MPCT}$ is as follows.

\textbf{Step 2:} Sparse representation of the CIR:
\begin{align*}
\boldsymbol{h}^{(\boldsymbol{x})}(n) = 
\begin{cases}
{\boldsymbol{\alpha}_\text{MPCT}(k)}~,&  \text{if}~ k^\text{th} \text{ MPC is present in bin}~n\\
0~, & {\rm otherwise}~
\end{cases}~,
\end{align*}
where $\boldsymbol{x} \in \{\boldsymbol{\tau}, \boldsymbol{\psi}, \boldsymbol{\theta}\}$. Here, $\boldsymbol{h}^{(\boldsymbol{x})} = [\boldsymbol{h}^{(\boldsymbol{x})}(1),...,\boldsymbol{h}^{(\boldsymbol{x})}(N)]$ represents the CIR in the domain of $\boldsymbol{x} $, and is a sparse vector whose length depends on the domain of $\boldsymbol{x}$ and the resolution. For instance, if $\boldsymbol{x} = \boldsymbol{\tau}$, then $\boldsymbol{h}^{(\boldsymbol{\tau})}$ corresponds to the CIR in TOA domain whose length depends on the time resolution and time range considered. In our simulations, we capture the TOA terms in the range of 1 nanoseconds to 1000~ns with a resolution of 1~ns. Thus, $\boldsymbol{h}^{(\boldsymbol{\tau})}$ is a vector of size 1000 capturing the AnP term with respect to the TOA.  Similarly, when $\boldsymbol{x} = \psi$ it corresponds to the AOD term and $N$ depends on the resolution of elevation and azimuth angle. In our simulations, we choose the angular resolution of 1~degree. Based on the above notation, $\boldsymbol{h}_{i}^{(\boldsymbol{x})}$ is the RV $\boldsymbol{h}^{(\boldsymbol{x})}$ at frequency band $f_i$. 

In order to assess the amount of correlation for a given RV across different bands, we use the Pearson correlation coefficient formula which is given by
\begin{equation}\small
\Psi_{\boldsymbol{h}^{(x)}_{i},\boldsymbol{h}^{(x)}_{j}} \hspace{-1mm}= \hspace{-1mm}\frac{1}{N-1}\sum_{n=1}^N \left({\frac{{\boldsymbol{h}^{(x)}_{i}(n)} - \mu_{\boldsymbol{h}^{(x)}_i} }{\sigma_{\boldsymbol{h}^{(x)}_i}}}\right)^{*} \left({\frac{{\boldsymbol{h}^{(x)}_{j}(n)} - \mu_{\boldsymbol{h}^{(x)}_j} }{\sigma_{\boldsymbol{h}^{(x)}_j}}}\right)\nonumber
\end{equation}\normalsize
where $\mu_{\boldsymbol{h}^{(x)}_i}=\frac{1}{N}\sum_{n=1}^N \boldsymbol{h}^{(x)}_{i}(n)$ and $\sigma_{\boldsymbol{h}^{(x)}_i}= \frac{1}{N-1} \sum_{n=1}^N |\boldsymbol{h}^{(x)}_{i}(n) - \mu_{\boldsymbol{h}^{(x)}_i}|^2$ are the mean and standard deviation of RV $\boldsymbol{h}^{(x)}$ at frequency band $f_i$, and $(.)^{*}$ denotes conjugate operation. 
Similarly, $\mu_{\boldsymbol{h}^{(x)}_j}$ and ${\sigma_{\boldsymbol{h}^{(x)}_j}}$ are the mean and standard deviation of RV $\boldsymbol{h}^{(x)}$ at frequency band $f_j$. 


To make the notation and correlation experiment more clear to the reader we provide the following example. For instance, consider $\Psi_{\boldsymbol{h}^{(\tau)}_{1},\boldsymbol{h}^{(\tau)}_{8}}$ with $\boldsymbol{x}$ being the TOA term. This corresponds to the TOA correlation\footnote{We used Matlab's \textit{corrcoef} function to calculate correlation. The $\mu_{\boldsymbol{x}_{i}}$  and $\sigma_{\boldsymbol{x}_{i}}$ are taken care by the Matlab function.}  across frequency band $f_1$ and $f_8$ evaluated at a particular vehicle location. The modulus of $\Psi_{\boldsymbol{h}^{(\tau)}_{1},\boldsymbol{h}^{(\tau)}_{8}}$ provides us with the insight of how correlated the TOA terms are at $f_1$ and $f_8$ evaluated at a location of interest. Similar to this we calculate the correlation of TOA, AOA, and AOD terms at each location for different frequencies and finally average over across all the locations in the LOS and NLOS trajectory separately. The averaged correlation for TOA, AOA, and AOD terms for both LOS and NLOS trajectory at different frequency bands are tabulated in Table~\ref{Correlation_25}. 

\subsection{Correlation Results}

The correlation matrix is normally symmetric since the correlation between $\boldsymbol{h}^{(x)}_{i}$ and $\boldsymbol{h}^{(x)}_{j}$ is the same as the correlation between $\boldsymbol{h}^{(x)}_{j}$ and $\boldsymbol{h}^{(x)}_{i}$. In order to save on the space, we merge the correlation results of the LOS and the NLOS trajectory in the same table: the upper triangular matrix corresponds to the correlation of the NLOS trajectory while the lower triangular matrix (in red) corresponds to the correlation of the LOS trajectory. Table~\ref{tab:T_DOD}(a) captures the correlation results of the TOA, while Table~\ref{tab:T_DOD}(b) and Table~\ref{tab:T_DOA}(c) correspond to the results of the azimuth angle of AOD and AOA, respectively. Results of elevation angle are omitted due to space constraint, but we note that the trend is similar to the azimuth angle.

\begin{table}[t!]
\centering
\captionsetup{justification=centering}
\caption{Correlation across different central frequencies for TOA, AOA, and AOD (MPCT$=40$~dB).}
\label{Correlation_25}
\hfill
\begin{subtable}[t]{0.48\textwidth}
\begin{tabular}{|P{0.2cm}|P{0.535cm}|P{0.535cm}|P{0.535cm}|P{0.535cm}|P{0.535cm}|P{0.535cm}|P{0.535cm}|P{0.535cm}|}

\hline
\textbf{\textit{f}}$_i$ & \textbf{\textit{f}}$_1$ &  \textbf{\textit{f}}$_2$ &  \textbf{\textit{f}}$_3$&  \textbf{\textit{f}}$_4$ &  \textbf{\textit{f}}$_5$ &  \textbf{\textit{f}}$_6$ &  \textbf{\textit{f}}$_7$ & \textbf{\textit{f}}$_8$\\
\hline
\textbf{\textit{f}}$_1$ & \textbf{1.00} & 0.83 & 0.84  &  0.81  &  0.81  &  0.65  &  0.56  &  0.45\\
\hline
\textbf{\textit{f}}$_2$ &  \textcolor{red}{0.90}  &  \textbf{1.00}  &  0.85  &  0.84    &0.83  &  0.65 &   0.56  &  0.45\\
\hline
\textbf{\textit{f}}$_3$ &  \textcolor{red}{0.90}  &   \textcolor{red}{0.93}   & \textbf{1.00}  &  0.82   & 0.84    &0.67   & 0.61    &0.45\\
\hline
\textbf{\textit{f}}$_4$ &  \textcolor{red}{0.93}  &   \textcolor{red}{0.93}  &   \textcolor{red}{0.93}   & \textbf{1.00}    &0.86    &{0.68}   & 0.64   & 0.48
\\
\hline
\textbf{\textit{f}}$_5$ &  \textcolor{red}{0.90}   &  \textcolor{red}{0.95}  &   \textcolor{red}{0.93}  &   \textcolor{red}{0.94}  &  \textbf{1.00}  &  {0.69}  &  0.65  &  0.60
\\
\hline
\textbf{\textit{f}}$_6$ &  \textcolor{red}{0.91}   &  \textcolor{red}{0.93}   &  \textcolor{red}{0.92}   &  \textcolor{red}{0.94}   &  \textcolor{red}{0.94}   & \textbf{1.00}   & 0.72   & 0.60
\\
\hline
\textbf{\textit{f}}$_7$ &  \textcolor{red}{0.91}   &  \textcolor{red}{0.93}  &   \textcolor{red}{0.94}  &   \textcolor{red}{0.95}   &  \textcolor{red}{0.94}   &  \textcolor{red}{0.95}   & \textbf{1.00}  & 0.62
\\
\hline
\textbf{\textit{f}}$_8$ &  \textcolor{red}{0.89}   &  \textcolor{red}{0.88}   &  \textcolor{red}{0.89}  &   \textcolor{red}{0.80}  &   \textcolor{red}{0.90}   &  \textcolor{red}{0.98}   &  \textcolor{red}{0.99}   & \textbf{1.00}\\
\hline
\end{tabular}
\caption{TOA.}
\end{subtable}\label{tab:T_TOA}

\hspace{\fill}
\begin{subtable}[t]{0.48\textwidth}
\begin{tabular}{|P{0.2cm}|P{0.535cm}|P{0.535cm}|P{0.535cm}|P{0.535cm}|P{0.535cm}|P{0.535cm}|P{0.535cm}|P{0.535cm}|}
\hline
\textbf{\textit{f}}$_i$ & \textbf{\textit{f}}$_1$ &  \textbf{\textit{f}}$_2$ &  \textbf{\textit{f}}$_3$&  \textbf{\textit{f}}$_4$ &  \textbf{\textit{f}}$_5$ &  \textbf{\textit{f}}$_6$ &  \textbf{\textit{f}}$_7$ & \textbf{\textit{f}}$_8$\\
\hline
\textbf{\textit{f}}$_1$  &\textbf{1.00}   & 0.57  & 0.56  &  0.48   & 0.47  &  0.46  &  0.46  &  0.45
\\
\hline
\textbf{\textit{f}}$_2$  & \textcolor{red}{0.86}    &\textbf{1.00}  & 0.57  &  0.53  & 0.53 &  0.48  &  0.58   & 0.48 
\\
\hline
\textbf{\textit{f}}$_3$ & \textcolor{red}{0.85}    & \textcolor{red}{0.90}   & \textbf{1.00} &  0.57  &  0.55   & 0.52   & 0.52   & 0.49
\\
\hline
\textbf{\textit{f}}$_4$ & \textcolor{red}{0.76}   &  \textcolor{red}{0.76}   &  \textcolor{red}{0.74}   & \textbf{1.00}   & 0.72   &{0.65}   & 0.65   & 0.62 
\\
\hline
\textbf{\textit{f}}$_5$ &  \textcolor{red}{0.77}   &  \textcolor{red}{0.77}    & \textcolor{red}{0.78}   &  \textcolor{red}{0.87}   & \textbf{1.00}  & {0.66}   & 0.65   & 0.62 
\\
\hline
\textbf{\textit{f}}$_6$ & \textcolor{red}{0.73}    & \textcolor{red}{0.74}   &  \textcolor{red}{0.77}   &  \textcolor{red}{0.84}   &  \textcolor{red}{0.84}    &\textbf{1.00}   &0.79   & 0.76 
\\
\hline
\textbf{\textit{f}}$_7$ & \textcolor{red}{0.75}    & \textcolor{red}{0.74}  &   \textcolor{red}{0.74}   &  \textcolor{red}{0.87}   &  \textcolor{red}{0.86}   &  \textcolor{red}{0.90}   & \textbf{1.00} &  0.80 
\\
\hline
\textbf{\textit{f}}$_8$  & \textcolor{red}{0.75}   &  \textcolor{red}{0.74}  &  \textcolor{red}{0.74}   &  \textcolor{red}{0.87}   &  \textcolor{red}{0.86}   &  \textcolor{red}{0.90}   &  \textcolor{red}{0.99}   & \textbf{1.00}
\\
\hline
\end{tabular}
\caption{AOD: Azimuth angle.}
\end{subtable}\label{tab:T_DOD}

\hspace{\fill}
\begin{subtable}[t]{0.48\textwidth}
\begin{tabular}{|P{0.2cm}|P{0.535cm}|P{0.535cm}|P{0.535cm}|P{0.535cm}|P{0.535cm}|P{0.535cm}|P{0.535cm}|P{0.535cm}|}
\hline
\textbf{\textit{f}}$_i$ & \textbf{\textit{f}}$_1$ &  \textbf{\textit{f}}$_2$ &  \textbf{\textit{f}}$_3$&  \textbf{\textit{f}}$_4$ &  \textbf{\textit{f}}$_5$ &  \textbf{\textit{f}}$_6$ &  \textbf{\textit{f}}$_7$ & \textbf{\textit{f}}$_8$\\
\hline
\textbf{\textit{f}}$_1$ &\textbf{1.00}  &0.74  &0.74  &  0.76 &   0.75  & 0.63&   0.55 &   0.47 \\
\hline
\textbf{\textit{f}}$_2$  &  \textcolor{red}{0.91}    &\textbf{1.00} &  0.78    &0.80  &  0.81   & 0.63   & 0.59   & 0.51 
\\
\hline
\textbf{\textit{f}}$_3$ & \textcolor{red}{0.91} &   \textcolor{red}{0.93} &   \textbf{1.00} &   0.79 &   0.77 &   0.68   & 0.59 &   0.55
\\
\hline
\textbf{\textit{f}}$_4$&  \textcolor{red}{0.93}  &   \textcolor{red}{0.94}   &  \textcolor{red}{0.95}  &  \textbf{1.00}   &0.79   & {{0.68}}    &0.61   & 0.54
\\
\hline
\textbf{\textit{f}}$_5$&    \textcolor{red}{0.91}    & \textcolor{red}{0.96}  &  \textcolor{red}{0.95}    & \textcolor{red}{0.95}   & \textbf{1.00}   & {0.65}   & 0.61    &0.53
\\
\hline
\textbf{\textit{f}}$_6$ &     \textcolor{red}{0.90}    & \textcolor{red}{0.94}     & \textcolor{red}{0.93} &  \textcolor{red}{0.94}   &  \textcolor{red}{0.94}     & \textbf{1.00}   &0.72   & 0.69
\\
\hline
\textbf{\textit{f}}$_7$ &    \textcolor{red}{0.90}   &  \textcolor{red}{0.93}   &  \textcolor{red}{0.93}  &   \textcolor{red}{0.95}  &   \textcolor{red}{0.94}   &  \textcolor{red}{0.96}   & \textbf{1.00} &  0.74
\\
\hline
\textbf{\textit{f}}$_8$ &     \textcolor{red}{0.90}   &  \textcolor{red}{0.93}   & \textcolor{red}{0.93}&  \textcolor{red}{0.93}   &  \textcolor{red}{0.95}   &  \textcolor{red}{0.94}   &  \textcolor{red}{0.96}    &\textbf{1.00}
\\
\hline
\end{tabular}
\caption{AOA: Azimuth angle.}
\end{subtable}\label{tab:T_DOA}\vspace{-0.5cm}
\end{table}

From Table \ref{Correlation_25}, it is evident that there exists a positive correlation for TOA, AOA, and AOD terms across all the frequency bands for both the LOS and the NLOS trajectory. In fact, the correlation is nearing 1 for LOS trajectory which implies that the terms of interest have a strong correlation. Thus, the information from $f_j$ can be used to make a more accurate prediction at $f_i$. On the other hand, for the NLOS trajectory, the correlation is moderate. This is because signals propagate poorly in NLOS conditions both in the sub-6 GHz and mmWave regime.

\vspace{-0.1cm}
\section{Channel Prediction with OOB measurements}\label{Simulation_Section}
In this section, we discuss whether/how we can predict the ``coarse CIR" at $f_i$  provided the estimates at $f_j$. From the previous section, we know that the temporal and angular characteristics of the LOS trajectory are highly correlated at different frequencies. Hence, we can make a more accurate prediction of TOA, AOA and AOD terms at $f_i$ provided measure at $f_j$ for the LOS trajectory. Thus, from here on, we restrict our discussion to the LOS trajectory. Without loss of generality, we consider predicting coarse CIR at $f_6$ = 28~GHz provided estimates at $f_5$ = 5.9~GHz. 
We split the problem of CIR prediction into 4 sub-problems: Finding the TOA term, the AOA term, the AOD term, and the AnP term at 28~GHz provided the estimates at 5.9~GHz.

\begin{figure}[t!]
\centerline{\includegraphics[width=1.1\columnwidth]{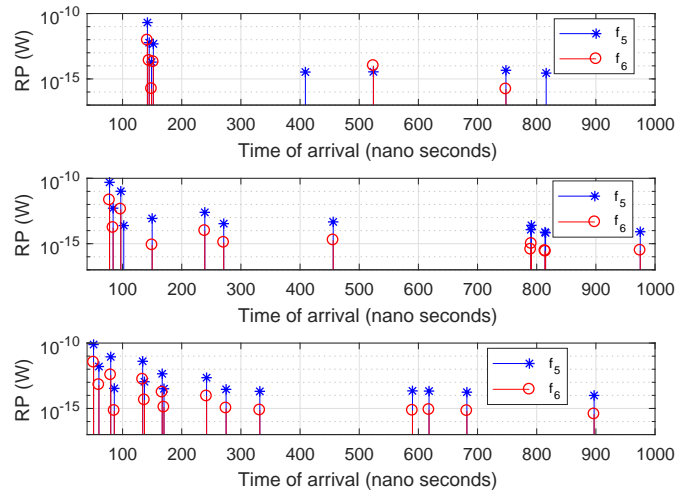}}
\vspace{-3mm}
\caption{Received power versus TOA at 5.9~GHz and 28~GHz. Top/middle/bottom: vehicle location 10/35/60.}\vspace{-3mm}
\label{TOA1}
\end{figure}

\subsubsection{TOA Term}
Fig. \ref{TOA1} shows the received power (RP) in the TOA domain with MPCT of 40 dB. The subplots correspond to CIR at 3 randomly picked vehicle locations (vehicle location 10, 35, and 60) in the LOS trajectory. It is evident from the plots that the TOA of the dominant paths at 5.9~GHz overlaps with the dominant paths at 28~GHz with a very high probability. In other words, there is a significant overlap, albeit not perfect. Due to the mismatch in the MPCs, we can only obtain a coarse CIR estimate and not perfect CIR. Based on Fig. \ref{TOA1}, we can say that the support set of the TOA terms are identical at 5.9~GHz and 28~GHz with a very high probability. 
Also, note the difference in the RP of the MPCs. This is obvious because the AnP behaves differently at different frequencies. That is at mmWave frequency, the RP is much lower due to diffuse scattering, and single/ multiple-bounce specular reflections relative to sub-6 GHz band.

\begin{figure}[h!]
\centerline{\includegraphics[scale=0.5]{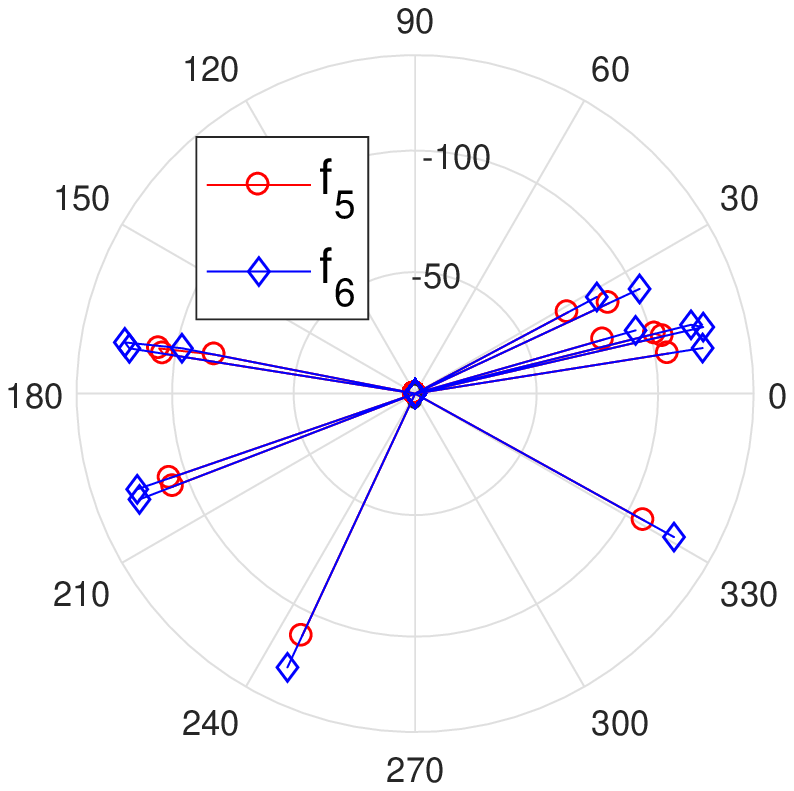}\hspace{2mm}
\includegraphics[scale =0.5]{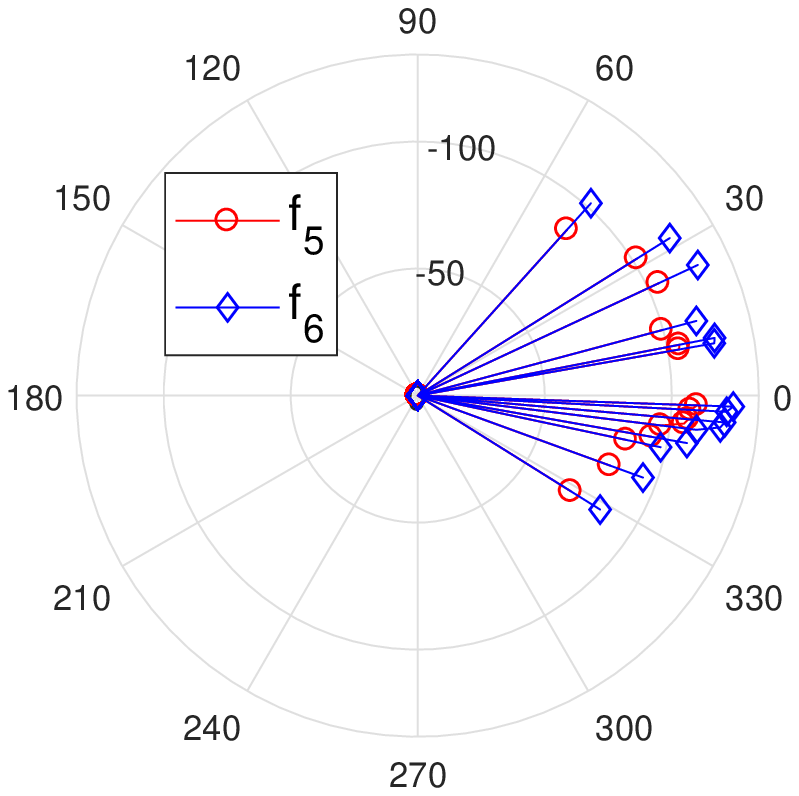}}
 \caption{AOA at 5.9~GHz and 28~GHz for the vehicle location 60 (LOS path); Left: azimuth angle, right: elevation angle.}
\label{DOA1}\vspace{-3mm}
\end{figure}

\begin{figure}[h!]
\centerline{\includegraphics[scale=0.5]{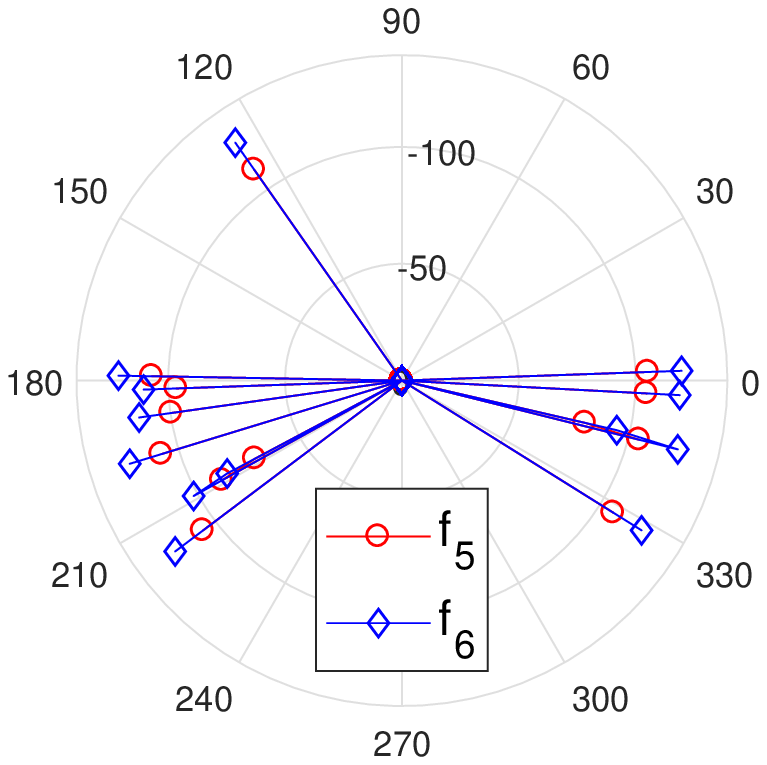}\hspace{2mm}
 \includegraphics[scale=0.5]{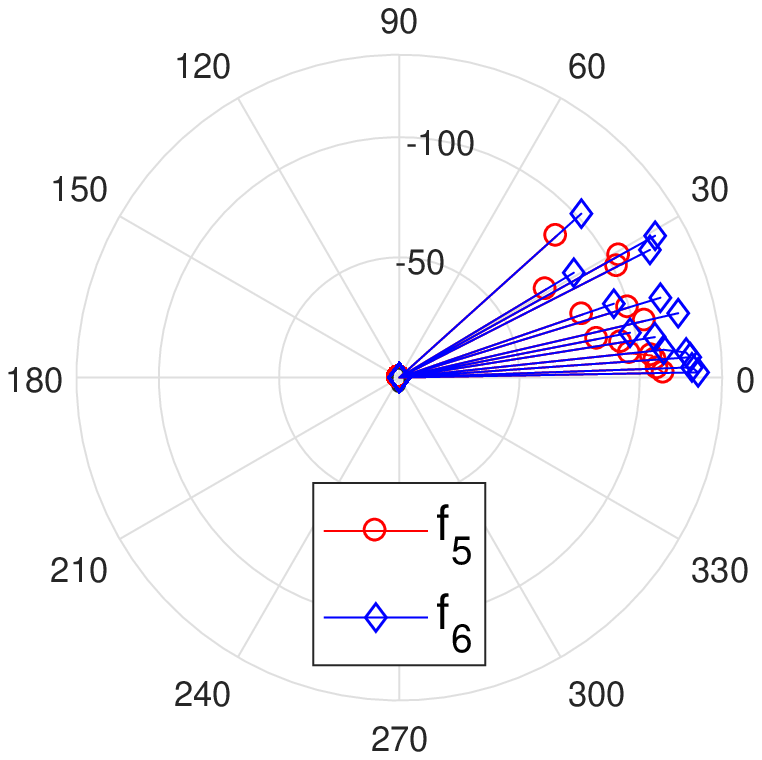}}
 \caption{AOD at 5.9 GHz and 28 GHz for the vehicle location 60 (LOS path); Left: azimuth angle, right: elevation angle.}
\label{DOD1}\vspace{-6mm}
\end{figure}

\subsubsection{AOA and AOD Terms}
Fig. \ref{DOA1} and Fig.~\ref{DOD1} plot both the elevation and azimuth angles of the AOA and the AOD in the polar domain, respectively. The radius in the circle corresponds to the RP of the MPCs in dBm. Note that the RP of all the MPCs is greater at 5.9~GHz relative to 28~GHz. Similar to the TOA result, the angles at which the MPCs arrive at the receiver and depart from the transmitter are almost the same at 5.9~GHz and 28~GHz. That is, the spatial characteristics are similar but not exact. Thus, the support set for the AOA and the AOD terms are identical at $f_5$ and $f_6$ with a very high probability. 


\subsubsection{AnP Term}
From the previous subsection we know that the support set of the TOA, AOA, AOD terms are the same with a very high probability. However, as obvious, the path loss (PL) terms are different. In other words, $\rho_k(f_5) \neq \rho_k(f_6)$ for all the considered MPCs~$k$. Often the PL terms can be taken care by the PL models which calculate the RP based on the distance dependency. For the sake of discussion, we assume close-in (CI) free space reference distance PL model \cite{PL_Model_CI} as:
\vspace{-0.25cm}
\begin{equation}\label{PL_Model}
PL(f_i, d) = FSPL(f_i,d_0) + 10 \gamma_{f_i} \log_{10}\left(\frac{d}{d_0} \right) + X_{\sigma}^{CI}\vspace{-3mm} 
\end{equation}
where $FSPL(f_i,d_0)=20\log_{10}\left(\frac{4\pi f_id_0}{c}\right)$ is the frequency dependent free space PL with $c$ being the speed of light, $d_0$ and $d$ being the reference distance for the antenna far-field and distance of UE from BS, respectively, $\gamma_{f_i}$ is the frequency dependent PL exponent term. We assume that the shadowing effect, $X_{\sigma}^{CI}$, is independent of frequency. Finding $\rho_k(f_6)$ provided $\rho_k(f_5)$, along with $\gamma_{f_5}$ and $\gamma_{f_6}$ is a simple substitution problem. That is, we can find the difference in RP by a simple substitution given by $FSPL(f_6 - f_5,d_0) + 10 (\gamma_{f_6} - \gamma_{f_5}) \log_{10}\left(\frac{d}{d_0} \right)$. This the difference in RP between the two bands of interest. Note that the accuracy of the prediction depends on $\gamma_{f_i}$, and that the PL model should be capable of capturing losses such as atmospheric,  penetration, and reflection losses which can be predominant in mmWave bands.

The term $e^{-j \phi_k(f_i)}$ in (\ref{DCIR1}), on the other hand, characterizes the phase, where $\phi_k(f_i)$ can be written as:\vspace{-0.15cm}
\begin{equation}\label{Phase_term}
\phi_k(f_i) = 2\pi f_i \tau_k(t) - \phi_{D_k} - \phi_0,\vspace{-1mm}
\end{equation}
where $\phi_{D_k}= 2 \pi\frac{v}{\lambda_i} \int_t \cos(\theta_k (t))$ is the Doppler phase shift due to Doppler frequency. Here, $v$ is the velocity term and $\lambda_i$ is the wavelength of $f_i$. If the Doppler frequency is zero or it is negligible, obtaining $\phi_k(f_i)$ provided $\phi_k(f_j)$ is straightforward with some substitution effort.  

In summary, it is possible to accurately predict coarse CIR at $f_i$ provided estimates at $f_j$ for the LOS trajectory because of significant similarity in the temporal and spatial characteristics. Thus, using temporal and spatial characteristics from the sub-6~GHz is highly beneficial in improving the resiliency and reliability of mmWave communication in the vehicular scenario.
\vspace{-0.3cm}

\section{Conclusion}\vspace{-1mm}
We used ray tracing simulations to evaluate the joint multipath channel statistics and the correlation among sub-6 GHz and mmWave bands for a vehicular trajectory in an urban environment. Our results show that for MPCT of 40 dB, the temporal/angular correlation among 5.9 GHz and 28 GHz bands for LOS (NLOS) path range between 0.84 to 0.94 (0.64 to 0.69). The high correlation in the LOS channel characteristics can be utilized for predicting temporal and angular characteristics in mmWave bands for LOS path and improving the resiliency and reliability of mmWave communications for vehicular scenarios. Our future work includes explicitly implementing beamforming and beam steering to evaluate correlation behavior and considering more complicated environments with multiple vehicles and mmWave access points. We will also explore conducting channel propagation experiments at multiple bands to verify that our observations from RT simulations are accurate in practice.
\vspace{-0.175cm}

\section*{Acknowledgement}
This research was supported in part by the U.S. National Science Foundation under the grant number ACI-1541108.

\vspace{-0.55cm}
\bibliographystyle{IEEEtran}\
\bibliography{bib_ICC1}




\end{document}